
\magnification=\magstep1
\overfullrule=0pt
\hsize=15.4truecm
{\nopagenumbers \line{\hfil UQAM-PHE-95/04}
\vskip1cm
\centerline{\bf NEUTRALINO CONTRIBUTION TO THE MASS DIFFERENCE OF}
\centerline{\bf $B_d^0-\overline B_d^0$}
\vskip2cm
\centerline{{G. COUTURE AND H. K\"ONIG}
\footnote*{email:couture@mercure.phy.uqam.ca, konig@osiris.phy.uqam.ca}}
\centerline{D\'epartement de Physique}
\centerline{Universit\'e du Qu\'ebec \`a Montr\'eal}
\centerline{C.P. 8888, Succ. Centre Ville, Montr\'eal}
\centerline{Qu\'ebec, Canada H3C 3P8}
\vskip2cm
\centerline{\bf ABSTRACT}\vskip.2cm\indent
\noindent
We present a detailed and complete calculation of the
neutralino contribution to the mass difference
$\Delta m_{B_d^0}/m_{B_d^0}$\ within the MSSM.
We include the complete mixing matrices of the neutralinos
and of the scalar partners of the left and right
handed bottom quark.
We find that the neutralino contribution is generally small but
can be of the same order of
the chargino contribution and much larger than that of the
gluino over a small range of $m_S$,
given $m_{g_2}$ and $\mu$ and for $\tan\beta\sim 50$.
\vskip2cm
\centerline{ November 1995}
\vfill\break}
\pageno=1
\noindent
{\bf I. INTRODUCTION}\hfill\break\vskip.2cm\noindent
In a recent paper [1] we presented a detailed analysis
of the charginos and the scalar partners of the up
quarks as well as of the gluino and scalar partners of
the down quarks contribution to the mass difference
$\Delta m_{B_d^0}/m_{B_d^0}$. We included the
mixing of the charginos and of the third
generation of the scalar partners of the up and
down quarks. It was shown that for large values
of the gluino mass ($m_{\tilde g}>100$\ GeV) and
of the soft supersymmetric breaking scalar mass ($m_S> 300$\ GeV)
the charginos give the most important contribution
to the mass difference in the $B$\ system, although they
couple only weakly to quarks and scalar quarks, whereas
the gluino couples strongly. Therefore we conlcuded
that it was not legitimate to neglect the contribution
of the neutralinos and the scalar partners of the down quarks
in front of the gluino and the scalar partners of
the down quarks, as is usually done in the literature, since the neutralinos
and charginos couple with same strength.
\hfill\break\indent
In this paper we present a detailed analysis of
contribution of the neutralinos
to the mass difference
$\Delta m_{B_d^0}/m_{B_d^0}$.
In our calculation we include the mixing
of the neutralinos and of the
scalar partners of left and right handed bottom
quark including one loop corrections.
Although the bottom quark mass is relatively small
the mixing might become important
if $\tan\beta=v_2/v_1\gg 1$\ where $v_{1,2}$\ are the
vacuum expectation values (vev's) of the Higgs particles
in the MSSM.
\hfill\break\indent
In the next section we present the calculation and discuss the
results in the third section. We end with the conclusions.
\hfill\break\vskip.2cm\noindent
{\bf II. NEUTRALINO CONTRIBUTION TO THE $B_d^0$\ SYSTEM}\vskip.2cm
For the complete analysis of the contribution of
the charginos and scalar up quarks and of the gluino
and scalar down quarks we refer the reader to [1].
Here we repeat only shortly the results of the standard
model (SM) contribution to $\Delta m_{B_d^0}/m_{B_d^0}$,
which is obtained by calculating the box diagrams, where W bosons
and up quarks are taken within the loop. After summation
over all quarks it turns out that the top quark gives the
main contribution to $\Delta m_{B_d^0}$. The result is
well known and given by [2]:
$$\eqalignno{{{\Delta m_{B_d^0}}\over{m_{B_d^0}}}=&
{{G_F^2}\over{6\pi^2}}f^2_BB_B\eta_tm^2_W
(K_{31}^\ast K_{33})^2S(x_t)&(1)\cr
S(x_t)=&x_t\lbrace {1\over 4}+{9\over 4}(1-x_t)^{-1}
-{3\over2}(1-x_t)^{-2}\rbrace -{3\over 2}{{{x_t}^3}\over
{(1-x_t)^3}}\ln x_t\cr}$$
$f_B$, $B_B$\ are the structure constant and the
Bag factor obtained by QCD sum rules and $\eta_t$\
a QCD correction factor [2].
For a large top quark mass their values are given by
$f_B=0.180$\ GeV, $B_B=1.17$\ and $\eta_t=0.55$ [3].
The experimental value is given by
$\Delta m_{B_d^0}/m_{B_d^0}\approx 6.4\times 10^{-14}$\
[4]
\hfill\break\indent
To obtain $\Delta m_{B_d^0}$\ with neutralinos
and scalar down quarks within the loop we have to
calculate all those diagrams as shown in Fig.1, where we
also include the so called "mass insertion" diagram
denoted with $\otimes$\ [5]. This notation means that in
this diagram $P_R(\rlap/ k+m)P_R=mP_R$\ remains whereas
the other one gives $P_R(\rlap/ k+m)P_L=\rlap/ kP_L$\
($P_{L,R}$\ are the projection operators).
In our calculation we take the full set of couplings as
given in Fig.24 of [6].
Furthermore as mentioned above
we include the mixing of the neutralinos and
the scalar partners of the left and right handed
down quarks.
For a more detailed description of the mixing of the charginos
and of the scalar top quarks we refer to [1,7]. Here we only
repeat the mixing of the scalar partner of the left and
right handed bottom quark,
that is instead of the current eigenstates
$\tilde q_{L,R}$\ we
work with the mass eigenstates
$$\tilde q_1=cos\Theta_q\tilde q_L+\sin\Theta_q\tilde q_R\qquad
\tilde q_2=-\sin\Theta_q\tilde q_L+\cos\Theta_q\tilde q_R\eqno(1)$$
A while ago it was shown in [8--9], that when
including loop effects flavour changing couplings
of the scalar partner of the left handed down quarks with
the gluinos are created, whereas the couplings of the
gluinos with the scalar partner of the right handed down
quarks remain flavour diagonal, that is only the scalar
partners of the left handed
down quarks have to be considered in the
relevant loop diagrams to $B_d^0-\overline B_d^0$\
mixing. When neutralinos are taken in the loop we
have to consider both couplings of the neutralinos
to the scalar partners of the left and right handed
down quarks.
For the mass matrix of the scalar bottom quark
we have to take at 1 loop level\footnote{$^1$}{To be more
exact the term $\vert c\vert m^2_t$\ should be read as
$\vert c\vert K^\dagger m_um_u^\dagger K$\ where $m_u$\ is
the diagonalized up quark matrix and $K$\ the KM-matrix.
But since $m_{u,d}^2\ll m^2_t$\ only the top quark is of
importance.}:
$$M^2_{\tilde b}=\left(\matrix{m_{\tilde b_L}^2+
m_b^2-0.42D_Z-\vert c\vert m_t^2
&-m_b(A_b+\mu\tan\beta)\cr
-m_b(A_b+\mu\tan\beta)&m_{\tilde b_R}^2+m_b^2-0.08D_Z\cr
}\right)\eqno(2)$$
with $T_3^b-e_b\sin\Theta_W=-0.42$\ and $e_b\sin\Theta_W=-0.08$.
$b$\ is understood as subscript for all three generations
down, strange and bottom.
$m_{\tilde b_{L,R}}$\ are soft SUSY breaking mass terms, $A_b$\
the parameter from the trilinear scalar interaction and $\mu$\
the mixing mass term of the Higgs bosons (in the analysis here
we take $m_{\tilde b_L}=m_S=m_{\tilde b_R}=A_b$).
The model dependent parameter $c$\
plays a crucial role in the calculation
of the gluino and scalar down quark contribution to the
mass difference in the $B_d^0$\ system. The value of $c$\
is negative and of order 1 ($\vert c\vert$ increases
with the soft SUSY breaking mass term $m_S$\ and decreases
with the top quark mass [10]). In the following we take
$c=-1$, although we keep in mind that it is more likely smaller in
magnitude.
The mixing term might only get important in the case
$\tan\beta\gg 1$.
\hfill\break\indent
For the mass matrix of the neutralinos we take eq.(A.19)
in [6]:
$$M_{\tilde N}=\left(\matrix{m_{g_1}&0&-m_Z\sin\theta_W
\cos\beta&m_Z\sin\theta_W\sin\beta\cr
0&m_{g_2}&m_Z\cos\theta_W\cos\beta&-m_Z\cos\theta_W
\sin\beta\cr-m_Z\sin\theta_W\cos\beta&m_Z\cos\theta_W\cos\beta&
0&-\mu\cr m_Z\sin\theta_W\sin\beta&-m_Z\cos\theta_W\sin\beta&
-\mu&0}\right)\eqno(3)$$
In order to have fewer free parameters
we also use the well known GUT relation between the
$U(1)$\ and $SU(2)$\ gaugino masses $m_{g_1}={5\over 3}
m_{g_2}\tan^2\Theta_W$
\footnote{$^2$}{There is also a similiar relation between
the gluino and $SU(2)$\ gaugino mass $m_{\tilde g}=(g_s/g_2)^2
m_{g_2}$}
We calculate the mass eigenvalues
and mixing angles numerically and take the smallest
eigenvalue to be higher than about $30$\ GeV [11].
\hfill\break\indent
After a lengthy but straightforward calculation we obtain
the following result when neutralinos and scalar down-type quarks
are running on the loop:
$$\eqalignno{{{\Delta m_{B_d^0}}\over{m_{B_d^0}}}=&
{{G_F^2}\over{(4\pi)^2}}f^2_BB_Bm^4_Z(K^\ast_{31}K_{33})^2
\lbrack Z^{\tilde N}_{11}-2 Z'^{\tilde N}_{31}+
\tilde Z^{\tilde N}_{33}\rbrack&(4)\cr
Z^{\tilde N}_{11}=&\sum\limits_{i,j=1,4}(T^L_iT^L_j-
T^R_iT^R_j)^2 G^{ij}_{\tilde d\tilde d}
+2T^{m_b}_iT^L_j(T_i^{m_b}T_j^R-T_j^{m_b}T_i^R)
\tilde F^{ij}_{\tilde d\tilde d}\cr
&+\lbrace T_i^{m_b}T^L_j(T_j^{m_b}T_i^L-T_i^{m_b}T_j^L)
+T_i^{m_b}T^R_j(T_j^{m_b}T_i^R-T_i^{m_b}T_j^R)\rbrace
2 _M\tilde F^{ij}_{\tilde d\tilde d}\cr
Z'^{\tilde N}_{31}=&\sum\limits_{i,j=1,4}\Bigl\lbrace
T^L_i T^L_j(T_i^LT_j^L-T_i^RT_j^R)
\lbrack c^2_{\Theta_b}G^{ij}_{\tilde b_1
\tilde d}+s^2_{\Theta_b}G^{ij}_{\tilde b_2\tilde d}\rbrack
\cr &+T^R_iT^R_j(T_i^RT_j^R-T_i^LT_j^L)
\lbrack s^2_{\Theta_b}G^{ij}_{\tilde b_1\tilde d}
+c^2_{\Theta_b}G^{ij}_{\tilde b_2\tilde d}\rbrack\cr
&+T_i^{m_b}T_j^L(T_i^{m_b}T_j^R-T_j^{m_b}T_i^R)
\lbrack c^2_{\Theta_b}\tilde F^{ij}_{\tilde b_1\tilde d}
+s^2_{\Theta_b}\tilde F^{ij}_{\tilde b_2\tilde d}\rbrack\cr
&+T_i^{m_b}T_j^L(T_j^{m_b}T_i^L-T_i^{m_b}T_j^L)
\lbrack c^2_{\Theta_b}2 _M\tilde F^{ij}_{\tilde b_1\tilde d}+
s^2_{\Theta_b}2 _M\tilde F^{ij}_{\tilde b_2\tilde d}\rbrack\cr
&+T_i^{m_b}T_j^R(T_i^{m_b}T_j^L-T_j^{m_b}T_i^L)
\lbrack s^2_{\Theta_b}\tilde F^{ij}_{\tilde b_1\tilde d}
+c^2_{\Theta_b}\tilde F^{ij}_{\tilde b_2\tilde d}\rbrack\cr
&+T_i^{m_b}T_j^R(T_j^{m_b}T_i^R-T_i^{m_b}T_j^R)
\lbrack s^2_{\Theta_b}2 _M\tilde F^{ij}_{\tilde b_1\tilde d}+
c^2_{\Theta_b}2 _M\tilde F^{ij}_{\tilde b_2\tilde d}\rbrack
\Bigr\rbrace\cr
\tilde Z^{\tilde N}_{33}=&\sum\limits_{i,j=1,4}\Bigl\lbrace
T_i^{L2} T_j^{L2}
\lbrack c^4_{\Theta_b}G^{ij}_{\tilde b_1
\tilde b_1}+2c^2_{\Theta_b}s^2_{\Theta_b}G^{ij}_{\tilde b_1
\tilde b_2}+s^4_{\Theta_b}G^{ij}_{\tilde b_2\tilde b_2}\rbrack\cr
&-2T^L_iT_j^LT^R_iT_j^R
\lbrack c^2_{\Theta_b}s^2_{\Theta_b}(G^{ij}_{\tilde b_1
\tilde b_1}+G^{ij}_{\tilde b_2\tilde b_2})+(c^4_{\Theta_b}
+s^4_{\Theta_b})G^{ij}_{\tilde b_1\tilde b_2}\rbrack\cr
&+T_i^{R2} T_j^{R2}
\lbrack s^4_{\Theta_b}G^{ij}_{\tilde b_1\tilde b_1}
+2c^2_{\Theta_b}s^2_{\Theta_b}G^{ij}_{\tilde b_1\tilde b_2}
+c^4_{\Theta_b}G^{ij}_{\tilde b_2\tilde b_2}\rbrack\cr
&+2T^{m_b}_iT^L_j(T_i^{m_b}T_j^R-T_j^{m_b}T_i^R)
\lbrack c^2_{\Theta_b}s^2_{\Theta_b}(\tilde F^{ij}_{\tilde b_1
\tilde b_1}+\tilde F^{ij}_{\tilde b_2\tilde b_2})+
(c^4_{\Theta_b}+s^4_{\Theta_b})\tilde F^{ij}_{\tilde b_1
\tilde b_2}\rbrack\cr
&+T_i^{m_b}T^L_j(T_j^{m_b}T_i^L-T_i^{m_b}T_j^L)
\lbrack c^4_{\Theta_b}2 _M\tilde F^{ij}_{\tilde b_1\tilde b_1}
+2c^2_{\Theta_b}s^2_{\Theta_b}2 _M\tilde F^{ij}_{\tilde b_1
\tilde b_2}+s^4_{\Theta_b}2 _M\tilde F^{ij}_{\tilde b_2\tilde b_2}
\rbrack\cr
&+T_i^{m_b}T^R_j(T_j^{m_b}T_i^R-T_i^{m_b}T_j^R)
\lbrack s^4_{\Theta_b}2 _M\tilde F^{ij}_{\tilde b_1\tilde b_1}
+2c^2_{\Theta_b}s^2_{\Theta_b}2 _M\tilde F^{ij}_{\tilde b_1
\tilde b_2}+c^4_{\Theta_b}2 _M\tilde F^{ij}_{\tilde b_2\tilde b_2}
\rbrack
\Bigr\rbrace\cr
G^{ij}_{ab}:=&\tilde F^{ij}_{ab}+2 _M\tilde F^{ij}_{ab}\cr
T^{m_b}_i=&{{m_b}\over{m_Z\cos\beta}}N_{i3}\cr
T_i^L=&e_d\sin2\Theta_W N'_{i1}-(1+2e_d\sin^2\Theta_W)N'_{i2}\cr
T_i^R=&-\lbrace e_d\sin2\Theta_W N'_{i1}-2e_d\sin^2\Theta_W
N'_{i2}\rbrace\cr
}$$
$c^2_{\Theta_b}=\cos^2\Theta_b$, $s^2_{\Theta_b}=\sin^2\Theta_b$\
and $\cos\beta$\ can be extracted from $\tan\beta$.
$N_{ij}$\ and $N'_{ij}$\ are the diagonalizing angles
as defined in eq.(A.20)\ and eq.(A.23) in [6] and taken to
be real. We calculate them numerically.
$\tilde F^{ij}_{ab}$\
and $_M\tilde F^{ij}_{ab}$\
are given in the appendix A.
$m_{i,j}=m_{\tilde N_{i,j}}$\ are the mass eigenvalues
of the neutralinos
and $m_{\tilde d, \tilde b_{1,2}}$\
the masses of the scalar down quark and the eigenstates
of the scalar bottom quark including the mixing and 1 loop
corrections.
$\tilde F^{ii}_{ab}$\ and $_M\tilde F^{ii}_{ab}$\
are the same functions as given in eq.C.2 in [12].
\hfill\break\indent
Since we neglected all quark masses
except that of the bottom
quark, we made use of $Z_{11}=Z_{12}=Z_{21}=Z_{22}$\
and $Z_{13}=Z_{31}=Z_{32}=Z_{23}$.
As was shown in [13]\ the mixing angles of the scalar
down quarks might also become important in the second
generation but it is safe to neglect them here since
$m_{\tilde s_1}\simeq m_{\tilde s_2}$\ and
since the mixing of the scalar quark is proportional
to the quark masses $c^2_{\Theta_b}=\cos^2\Theta_b\approx
1$, only for large values of $\tan\beta$\ the mixing angle
of the scalar bottom quark mass becomes more important.
\hfill\break\indent
\vskip.2cm\noindent
{\bf III. DISCUSSIONS}\vskip.2cm
We now present those contributions  for different values of gaugino
and scalar down quark masses.
We also vary $\tan\beta$ and the symmetry-breaking scales.
 As input parameter we take
$m_{\rm top}=174$\ GeV, $m_{\rm b} = 4.5$\ GeV, $\alpha = 1/137$\ and
for the strong coupling
constant $\alpha_s=0.1134$.
For a top quark mass of 174 GeV the SM result eq.1 gives
a value of $4.67\times 10^{-16}$.
\hfill\break\indent
In Figs. 2, we show the neutralino contribution and
compare them with the chargino contribution.
\hfill\break\indent
The global
behaviour is clear: for small values of $\tan\beta$ ($\sim 20$ or less) the
neutralino contribution is small compared to that of the chargino. On the
other hand, when $\tan\beta\sim 50$,
\footnote{$^3$}{Such high values for $\tan\beta$\ are {\bf preferred}
in models, which require the Yukawa couplings $h_t,\ h_b\
{\rm and}\ h_\tau$\ to meet at one point at the unification scale [14]}  the
neutralino contribution can be much
larger than that of the chargino for the smallest possible values of
$m_S$.  Unfortunately, as we can see on the figure, this contribution falls
very quickly as $m_S$ increases and becomes negligible as soon as $m_S$ is a
few tens of GeV's above its minimal value
(for smaller values of $m_S$\ the square of one of the mass eigenvalues of the
scalar bottom quark becomes negative).
\hfill\break\indent
A few comments are in order concerning the chargino contribution. First,
we note that with the relation $m_{\tilde g}=
(g_s/g_2)^2m_{g_2}$\ the gluino mass is in the order of TeV
and therefore negligible compared
to the chargino contribution as we have shown in [1].
Second, if we allow this mass to
be arbitrary, the gluino contribution can be dominant for small
gluino mass ($\sim 200$ GeV) and small values of $m_S$ ($\sim 300$ GeV or
less). Even in the best cases, this contribution becomes negligible for values
of $m_S$ of 500 GeV or less; even for $\tan\beta\sim 50$.
\hfill\break\indent
We note that values of
$\mu$ like 100 GeV or 400 GeV for a value of $m_{g_2}$ of 200 GeV will increase
(in magnitude) the contributions from the charginos
while negative values
of $\mu$ will decrease them in magnitude. Furthermore, the effects of the
mixing of the scalar partners of the top and bottom quarks are more
important for
large values of $m_S$: the contributions from the charginos
and neutralinos do not decrease as
quickly with the mixing. For small values of $m_S$, there is also an
enhancement.
\hfill\break\indent
In Fig.2 in [1] we also have presented the contribution
of the charged Higgs and shown that it is quite large
for small Higgs masses, but decreases rapidly for values
of $\tan\beta$\ larger than $1$. In [1] we also
discussed the importance of the
 factor $c$ that enters the b-quark mixing matrix and
shown that reducing it from 1 to 1/2 reduces the gluino contributions by
a factor of $\sim 8$ for small values of $m_S~(\sim 200~GeV)$ while
it reduces them by a factor of $\sim 4$ for very large values of
$m_S~(\sim 1~TeV)$. The first factor will vary with $m_{g_2}$, $\mu$ and
$\tan\beta$ but the
reduction of 4 for large $m_S$ is rather stable. Typically it will vary between
3.5 and 4.1. It can be understood by the fact that for those large scales, the
mass eigenvalues are almost insensitive to $c$ but the mixing angles are almost
linearly dependant on $c$. The same happens for the neutralino
contribution.
\hfill\break\indent
Finally, one must not forget that in eq.(4)
$K_{31}^\ast K_{33}$\ have not necessarily the same
values as in the SM.
This was shown in [8]: the
Kobayashi--Maskawa matrix in the couplings of the
neutralinos to quarks and scalar quarks is multiplied by
another matrix $V_d$, which can be parametrized as follows:
$$V_d=\left(\matrix{1&\varepsilon_d&\varepsilon_d^2\cr
 -\varepsilon_d&1&\varepsilon_d\cr-
\varepsilon_d^2&-\varepsilon_d&1\cr}\right)
\eqno (5)$$
so that $K=V_d$\footnote{$^4$}{For simplicity we
have taken the flavour non diagonal couplings of the neutralinos
to the scalar partners of the left and right handed down quarks
to be of the same order}.
For $\varepsilon_d=0.1$\ $K^\ast_{31}K_{33}$\
is identical to the SM values, whereas $\varepsilon_d=0.5$\
enhances it by a factor of $25$\ and $\varepsilon=0.3$\ by $9$.
Considering that these values are to be squared in the
mass difference of the $B^0_D$\ system we can use that
enhancement to put limits on $\varepsilon_d$.
\hfill\break\indent
In the case at hand
$\varepsilon_d$\ has to be smaller than $0.1$\ to keep the results
lower than the measured value of $\Delta m_{B^0_d}/m_{B^0_d}$.
This is not very constraining yet but it is already better than the limit one
can get from current data on rare Kaon decays [7].
\hfill\break\vskip.12cm\noindent
{\bf IV. CONCLUSIONS}\vskip.12cm
In this paper we presented the
contributions from
neutralinos and scalar down quarks
to the mass difference in the $B_d^0$\
system via box diagrams.
We gave exact results
and included the mixing of the neutralinos and
the mixing of the scalar bottom quark.
We find that this contribution is in general small but can be important
for large values of $\tan \beta$ ($ \sim 50$) and the smallest possible values
of $m_S$, given $m_{g_2}$ and $\mu$. As soon as $m_S$ is a few tens of GeV's
above its minimum values, this contribution is very small.
\hfill\break\indent
With this paper we complete our
analysis within the MSSM where all its particles
are taken within the relevant box diagram without
neglecting any mixing angles and mass eigenvalues.
We conclude that, generally the most important
contribution is that of the chargino. Over a narrow range of $m_S$, for
given values of $m_{g_2}$ and $\mu$ and for $\tan \beta \sim 50$,
the neutralino contribution can be quite large and dominant.
The gluino contribution is
negligible for gluino masses larger than $\sim 300$ GeV. All these are small
compared to the charged Higgs contribution for small values of $\tan \beta$,
but
this one is negligible ($\Delta^{Higgs}/\Delta^{SM}\sim 1.2\times 10^{-4}$) for
$\tan \beta = 50$\ and a charged Higgs mass of $100$\ GeV.
We hope that these results will
find some application with the upcoming {\it B-factories}.
\hfill\break\vskip.1cm\noindent
{\bf V. ACKNOWLEDGMENTS}\vskip.12cm
\noindent
We want to thank our colleague Cherif Hamzaoui for fruitful discussions.
One of us (H.K.) would like to thank the physics department
of Carleton university for the use of their computer facilities.
The figures were done with the
program PLOTDATA from TRIUMF and we used the CERN-Library to
diagonalize the neutralino mass matrix.
\hfill\break\indent
This work was partially funded by funds from the N.S.E.R.C. of
Canada and les Fonds F.C.A.R. du Qu\'ebec.
\hfill\break\vskip.12cm\noindent
{\bf VI. APPENDIX A}\vskip.12cm
For the box diagram we have to calculate the following
integrals:
$$\eqalignno{F^{\mu\nu}_{abij}:=&\int{{d^4k}\over{(2\pi)^4}}
{{k^\mu k^\nu}\over{(k^2-m_a^2)(k^2-m_b^2)(k^2-m_i^2)(k^2-m_j^2)}}
&(A0)\cr
F^{\mu\nu}_{abij}=:&{{+ig^{\mu\nu}}\over{4(4\pi)^2}}
\tilde F^{ij}_{ab}\cr}$$
$m^2_i\not=m^2_j\not=m^2_a\not=m^2_b$
$$\eqalignno{\tilde F^{ij}_{ab}=-{1\over{(m_j^2-m_i^2)(m_b^2-
m_a^2)}}\bigl\lbrace&{1\over{(m_i^2-m_a^2)(m_i^2-m_b^2)}}
\lbrack m_i^4(m_b^2\ln{{m_b^2}\over{m_i^2}}-m_a^2\ln {{m_a^2}
\over{m_i^2}})\cr
&-m^2_im^2_am_b^2\ln{{m_b^2}\over{m_a^2}}\rbrack-
(m^2_i\leftrightarrow m^2_j)\bigr\rbrace&(A1)\cr}$$
$$\eqalignno{m_j^2\not=&m_i^2\not=m^2_a=m^2_b\cr
\tilde F^{ij}_{aa}=&-{1\over{(m_j^2-m_i^2)}}\bigl\lbrace
{1\over{(m_i^2-m_a^2)}}\lbrack m^2_a-{{m_i^4}\over{(m_i^2-
m_a^2)}}\ln{{m_i^2}\over{m_a^2}}\rbrack-(m_i^2\leftrightarrow
m_j^2)\bigr\rbrace&(A2)\cr
m_i^2=&m_j^2\not=m_a^2\not=m^2_b\cr
\tilde F^{ii}_{ab}=&\tilde F^{ij}_{aa}(m_a^2\leftrightarrow
m_i^2,\ m_b^2\leftrightarrow m_j^2)&(A3)\cr
m_j^2=&m_i^2\not= m_a^2=m_b^2\cr
\tilde F^{ii}_{aa}=&-{{(m_i^2+m_a^2)}\over{(m_i^2-m_a^2)^2}}
\lbrace 1-{{2m_i^2m_a^2}\over{(m_i^4-m_a^4)}}\ln{{
m_i^2}\over{m_a^2}}\rbrace &(A4)\cr}$$
The second integral is given by:
$$\eqalignno{_MF^{ij}_{ab}:=&\int{{d^4k}\over{(2\pi)^4}}
{{m_im_j}\over{(k^2-m_a^2)(k^2-m_b^2)(k^2-m_i^2)(k^2-m_j^2)}}
&(A5)\cr
_MF^{ij}_{ab}=:&{{-ig^{\mu\nu}}\over{(4\pi)^2}} _M\tilde F^{ij}
_{ab}\cr}$$
$m^2_i\not=m^2_j\not=m^2_a\not=m^2_b$
$$\eqalignno{_M\tilde F^{ij}_{ab}=-{{m_im_j}\over{(m_j^2-m_i^2)(m_b^2-
m_a^2)}}\bigl\lbrace&{1\over{(m_i^2-m_a^2)(m_i^2-m_b^2)}}
\lbrack m_i^2(m_b^2\ln{{m_b^2}\over{m_i^2}}-m_a^2\ln {{m_a^2}
\over{m_i^2}})\cr
&-m^2_am_b^2\ln{{m_b^2}\over{m_a^2}}\rbrack-
(m^2_i\leftrightarrow m^2_j)\bigr\rbrace&(A6)\cr}$$
$$\eqalignno{m_j^2\not=&m_i^2\not=m^2_a=m^2_b\cr
_M\tilde F^{ij}_{aa}=&-{{m_im_j}\over
{(m_j^2-m_i^2)}}\bigl\lbrace
{1\over{(m_i^2-m_a^2)}}\lbrack 1-{{m_i^2}\over{(m_i^2-
m_a^2)}}\ln{{m_i^2}\over{m_a^2}}\rbrack-(m_i^2\leftrightarrow
m_j^2)\bigr\rbrace&(A7)\cr
m_i^2=&m_j^2\not=m_a^2\not=m^2_b\cr
_M\tilde F^{ii}_{ab}=&_M\tilde F^{ij}_{aa}(m_a^2\leftrightarrow
m_i^2,\ m_b^2\leftrightarrow m_j^2,\ m_im_j
\rightarrow m_i^2)&(A8)\cr
m_j^2=&m_i^2\not= m_a^2=m_b^2\cr
_M\tilde F^{ii}_{aa}=&-{{m_i^2}\over{(m_i^2-m_a^2)^2}}
\lbrace 2+{{(m_i^2+m_a^2)}\over{(m_i^2-m_a^2)}}\ln{{
m_a^2}\over{m_i^2}}\rbrace&(A9)\cr}$$
\hfill\break\vskip.12cm\noindent
{\bf REFERENCES}\vskip.12cm
\item{[\ 1]}G. Couture and H. K\"onig, "Reanalysis
of the mass difference of $\Delta m_{B_d^0}/m_{B_d^0}$\
within the minimal supersymmetric standard model",
hep-ph/9505315, to be published in Z.Phys.{\bf C}.
\item{[\ 2]} A.J. Buras, W. Slominski and H. Steger,
Nucl. Phys.{\bf B238}(1984)529, \hfill\break
 Nucl.Phys.{\bf B245}(1984)369
\item{[\ 3]}A.J. Buras, "Rare Deacys, CP Violation and
QCD", hep-ph/9503262.
\item{[\ 4]} Review of Particle Properties, part 1,
Phys. Rev. {\bf D50}(1994)1172.
\item{[\ 5]}J. Ellis and D.V. Nanopoulos, Phys.Lett.
{\bf 110B}(1982)44.
\item{[\ 6]}J.F. Gunion and H. Haber, Nucl.Phys.{\bf B272}
(1986)1.
\item{[\ 7]}G. Couture and H. K\"onig,
"$K^+\rightarrow \pi^+\nu\overline\nu$\
and $K^0_L\rightarrow\mu^+\mu^-$\ decays
within the minimal supersymmetric standard model",
to be published in Z.Phys. {\bf C}.
\item{[\ 8]}M.J. Duncan, Nucl.Phys.
{\bf B221}(1983)285.
\item{[\ 9]}J.F. Donoghue, H.P. Nilles and D. Wyler,
 Phys. Lett. {\bf 128B}
\item{[10]} S. Bertolini, F. Borzumati and A. Masiero,
Phys. Lett{\bf 194B}(1987)551. Erratum: ibid{\bf 198B}(1987)590.
\item{[11]}D. Decamp et al.:Phys.Lett.{\bf B244}(1990)541;
\hfil\hfil\break
\noindent M.Z.Akrawy et al.:Phys.Lett.{\bf B248}(1990)211.
\item{[12]}T. Inami and C.S. Lim, Progr. Theor. Phys.{\bf 65}
(1981)297.
\item{[13]}H. K\"onig, "The decay $H_2^0\rightarrow gg$\
within the minimal supersymetric standard model", hep-ph/9504433,
to be published in Z.Phys.{\bf C}.
\item{[14]}N. Polonsky, PhD thesis, hep-ph/9411378 and references
therein.
\hfill\break\vskip.12cm\noindent
{\bf FIGURE CAPTIONS}\vskip.12cm
\item{Fig.1}The box diagrams with scalar down quarks
and neutralinos
within the loop including the mass insertion diagram.
\item{Fig.2} The ratios
$\Delta m_{B^0_d}^{\rm Neutralino}/\Delta m_{B^0_d}^{\rm SM}$\ for
$tan\beta = 2,~5$ (dotted line, the two lines are on top of each other) and
$tan\beta = 50$ (dash)
and $\Delta m_{B^0_d}^{\rm Chargino}/\Delta m_{B^0_d}^{\rm SM}$\
for
$tan\beta = 2$ (very long dash-dot), $tan\beta = 5$ (long dash-dot), and
$tan\beta = 50$ (dash-dot) as a function of $m_S$ with
$m_{g_2} = \mu = 200~GeV$.
\hfil\hfil\break
\bye